\documentclass[11pt]{article}

\usepackage{amsmath,amssymb,amsfonts}
\usepackage{graphicx}
\usepackage{hyperref}

\begin{document}

\def\Tr{{\rm Tr }\ }

\newcommand{\ack}[1]{[{\bf Pfft!: {#1}}]}
\newcommand\bC{\mathbb{C}}
\newcommand\bP{\mathbb{P}}
\newcommand\bR{\mathbb{R}}
\newcommand\bZ{\mathbb{Z}}

\newcommand\dlangle{\langle\langle}
\newcommand\drangle{\rangle\rangle}
\newcommand{\bra}{\langle}
\newcommand{\ket}{\rangle}

\newcommand{\beq}{\begin{equation}}
\newcommand{\eeq}{\end{equation}}
\newcommand{\beqn}{\begin{eqnarray}}
\newcommand{\eeqn}{\end{eqnarray}}

\newcommand{\beql}[1]{\begin{equation}\label{eq:#1}}
\newcommand{\eq}[1]{(\ref{eq:#1})}
\newcommand{\eref}[1]{(\ref{#1})}

\newcommand{\pa}{\partial}

\def\del{\nabla}
\def\grad{\nabla}
\def\tr{\hbox{tr}}
\def\perp{\bot}
\def\half{\frac{1}{2}}

\renewcommand{\thepage}{\arabic{page}}
\setcounter{page}{1}

\rightline{hep-th/0602081} \rightline{ILL-(TH)-06-03} \rightline{HIP-2006-04/TH} \rightline{YITP-06-08}

\vskip 0.75 cm
\renewcommand{\thefootnote}{\fnsymbol{footnote}}
\centerline{\Large \bf The Rolling Tachyon } \centerline{\Large \bf Boundary Conformal Field Theory } \centerline{\Large \bf on an
Orbifold} 
\vskip 0.75 cm

\centerline{{\bf Shinsuke Kawai${}^{1}$\footnote{skawai@yukawa.kyoto-u.ac.jp},
Esko Keski-Vakkuri${}^{2}$\footnote{esko.keski-vakkuri@helsinki.fi},
}}
\centerline{{\bf Robert G. Leigh${}^{3}$\footnote{rgleigh@uiuc.edu} and
Sean Nowling${}^{3}$\footnote{nowling@students.uiuc.edu}
}}
\vskip .5cm
\centerline{${}^1$\it YITP, Kyoto University, Kyoto 606-8502, Japan}
\vskip .5cm
\centerline{${}^2$\it Helsinki Institute of Physics and Department of Physical Sciences,}
\centerline{\it P.O. Box 64, FIN-00014 University of Helsinki, Finland}
\vskip .5cm
\centerline{${}^3$\it Department of Physics,
University of Illinois at Urbana-Champaign}
\centerline{\it 1110 West Green Street, Urbana, IL 61801-3080, USA}
\vskip .5cm

\setcounter{footnote}{0}
\renewcommand{\thefootnote}{\arabic{footnote}}

\begin{abstract}
We consider the non-trivial boundary conformal field theory with exactly
marginal boundary deformation. In recent years this deformation has been
studied in the context of rolling tachyons and S-branes in string
theory. Here we study the problem directly from an open string point of
view, at one loop. We formulate the theory of the $\bZ_2$ reflection
orbifold.  To do so, we extend fermionization techniques originally
introduced by Polchinski and Thorlacius. We also explain how to perform
the open string computations at arbitrary (rational) radius, by
consistently constructing the corresponding shift orbifold, and show in
what sense these are related to known boundary states. In a companion
paper, we use these results in a cosmological context involving decaying
branes.
\end{abstract}

\newpage
\section{Introduction}\label{intro}

The $c=1$ conformal field theory on worldsheets with boundary is known
to have a boundary interaction
\[
-\frac{\lambda}{2}\int_{\partial\Sigma} ds\ e^{iX/\sqrt{\alpha'}}+h.c.
\]
which is exactly marginal. This theory was originally studied at
self-dual radius ($X\in S^1_{R=\sqrt{\alpha'}}$) by Callan et al
\cite{Callan}, where the marginality was established. Although there is
renormalization of $\lambda$ in perturbation theory, due to collisions
of $e^{iX/\sqrt{\alpha'}}$ with $e^{-iX/\sqrt{\alpha'}}$, it is possible
to absorb these divergences and obtain an RG-stationary coupling. At
self-dual radius, the operators of the theory organize themselves into
multiplets of an $\widehat{SU(2)}_1$ current algebra, and this structure
plays an important organizing r\^ole in the analysis.

A complementary analysis of this theory, at infinite radius, was later
provided by Polchinski and Thorlacius \cite{PT}. By introducing
auxiliary bosonic fields, it is possible to re-write the theory in terms
of free fermions with a quadratic boundary interaction. This essentially
constitutes a regularization of the theory (different than the one noted
above) and is tractable because the action is quadratic in fermions
(from the bulk worldsheet point of view, there are mass terms with
$\delta$-function support--these are both classically and quantum
mechanically marginal).

In this paper, we will consider the extension of this theory to other
backgrounds, including a $\bZ_2$ reflection orbifold, as well as circles
of arbitrary rational radius. We were led into this work by
considerations of S-brane solutions \cite{Gutperle:2002ai,Sen,Larsen} in Lorentzian
orbifold backgrounds \cite{BHKN,BKLNS}. In a companion paper
\cite{uscosmo} (see also \cite{KKLN}), we use the results of the present
paper; following a Wick rotation, the results are applicable to the
discussion of `fractional S-branes,' objects which may be of importance
in cosmological orbifold backgrounds.

Our calculations will be formulated and presented directly in the open
string channel. Passing these results to the closed string channel then
allows for the determination of boundary states\footnote{For related work
on boundary states on deformed boundary conformal field theory on the orbifold,
see also \cite{Recknagel:1998ih}.}. Away from self-dual
radius, we must formulate the theories of interest in the fermionic
picture.  In particular, this was originally formulated at infinite
radius. For the orbifold theory, it is necessary to carefully consider
various subtleties of the fermionic construction. As a result, we have
organized the paper as follows. In Section \ref{bdycft}, we set up
notation and discuss some standard boundary states of the undeformed
theory. Then, in Section \ref{Adsorp}, we review the standard bosonic
treatment of the bosonic theory at self-dual radius. In Section
\ref{fermionic} we then review the fermionic construction of the
infinite radius deformed theory. As we have mentioned above, there are a
number of subtleties involved in extending this analysis to the orbifold
theory, and thus we take the liberty of going into some detail in this
review. In this section we also review how the corresponding boundary
states can be written in terms of a projection operator acting on
$\widehat{SU(2)}_1$ Ishibashi states. We then discuss how finite radius
theories may be constructed in the fermionic picture. In constructing
these, there are both classical and quantum consistency conditions in
the fermionic path integral to which we must pay attention. Doing so
gives rise in the end to boundary states that can be written using
projection operators, and these boundary states transform in a simple
way under T-duality. In Section \ref{orbifoldsec}, we construct the
orbifold theory in the fermionic description. In so doing, we introduce
a number of consistency checks to ensure that the results are correct.

The boundary states, including contributions from both twisted and
untwisted sectors, may be constructed at arbitrary radius. In the
companion paper \cite{uscosmo}, we construct all of the fractional
boundary states, and show that they have a consistent interpretation. In
particular, we can see explicitly that the although the finite radius
theory has two orbifold fixed points, the infinite radius theory has but
one. This is reflected nicely in the structure of the boundary states.

\section{Boundary CFT of a free boson}\label{bdycft}

In order to set notation and collect some known results, we first consider the undeformed boundary conformal theory on a circle of
radius $R$. Free open bosonic string theory, with action $\frac{1}{4\pi\alpha'}\int d^2\sigma\sqrt{-h}h^{ab}\partial_a
X\partial_bX$, on a strip $\sigma\in (0,r), \tau\in (0,\ell)$, has mode expansion
\[
\hat X(\sigma, \tau) = \hat x+\frac{2\pi\alpha'}{r}\tau\hat p +i\sqrt{2\alpha'}\sum_{n\in\bZ\atop n\neq 0}\frac{1}{n}\hat\alpha_n
\cos \left(\frac{\pi n\sigma}{r}\right) e^{-i\pi n\tau/r}
\]
where we have assumed Neumann boundary conditions at $\sigma=0,r$. The
spectrum of $\hat p$ is determined by the compactification radius,
$spectrum(\hat p)=\bZ/R$. With this normalization, the vertex operator
$e^{ikX}$ has conformal dimension $\Delta=\alpha' k^2$.

If we compute the NN annulus amplitude (the open string partition
function with Neumann boundary conditions on each boundary), we may
obtain information on the Neumann boundary state of the closed string
channel. This is
\beq
\label{ANN}
{\cal A}_{NN} =\frac{1}{\eta(q)}\sum_n q^{\alpha' n^2/R^2} \ .
\eeq
We use the notation $q=e^{-\pi t}$, $t=\ell/r$.
This may be re-written as
\beq
{\cal A}_{NN} =\frac{R}{\eta(q)}\int dp\ q^{\alpha' p^2}\sum_m e^{2\pi ipRm} \ ,
\eeq
which can be understood as explicitly implementing the shift orbifold to
finite radius within the infinite radius theory. This form will be
important later.

In the present NN case, at finite radius, it is also possible to introduce a Wilson line, and we record the result \cite{AfflOsh}
here \beq \label{ANNWil} {\cal A}_{NN}(\Delta\theta) =\frac{1}{\eta(q)}\sum_n q^{\alpha' (n/R+\Delta\theta/2\pi R)^2} \ . \eeq By
Poisson resummation, with notation $\tilde q=e^{-2\pi /t}$, we find \beq\label{eq:clchanNN} {\cal A}_{NN}(\Delta\theta) =
\frac{R}{\sqrt{2\alpha'}\eta(\tilde q^2)} \sum_{m\in\bZ} (\tilde q^2)^{m^2R^2/4\alpha'} e^{-im\Delta\theta} \ . \eeq In this
channel, we identify boundary states via \cite{Cardy} \beq {\cal A}_{NN} (\Delta\theta) \equiv \langle N,\theta | \Delta(\tilde
q)|N,\theta+\Delta\theta\rangle \eeq with $\Delta(\tilde q)$ the closed string propagator. We may write the boundary state in
oscillator form as \cite{AfflOsh}
\[|N,\theta\rangle= 2^{-1/4} e^{\sum_k\alpha_k\tilde\alpha_k}|0\rangle_{Fock}\otimes
\sum_{n\in\bZ} e^{in\theta}|\frac{nR}{\sqrt{\alpha'}},-\frac{nR}{\sqrt{\alpha'}}\rangle
\]
In this form, it is clear that the Neumann boundary state has zero
momentum,\footnote{We record the Dirichlet boundary state at self-dual
radius  \[ |D,x\rangle\sim
2^{-1/4}e^{-\sum_k\alpha_k\tilde\alpha_k}|0\rangle_{Fock}\otimes\sum_n
e^{-inx/\sqrt{\alpha'}}|\frac{n}{\sqrt{\alpha'}},\frac{n}{\sqrt{\alpha'}}\rangle\]}
and is at fixed $\tilde X\equiv X_L-X_R$.

Note that at self-dual radius, $R=\sqrt{\alpha'}$, the conformal
dimensions are square integers. In fact, there is an $\widehat{SU(2)}$
current algebra that classifies the spectrum (see e.g. \cite{BYB}). In this case,
(\ref{eq:clchanNN}) can be rewritten \cite{GR}
\beq
{\cal A}_{NN}(\Delta \theta )
= \frac{1}{\sqrt{2}} \sum_{j=0,1/2,1,\ldots} \chi_{j^2}^{Vir}(\tilde q^2)\chi_j^{SU(2)}
(e^{-2i\Delta\theta J_0^3})
\eeq
with $SU(2)$ characters
\beq
\chi_j^{SU(2)}(g)=Tr_{j}{\cal D}^{(j)}(g) \ ,
\eeq
where ${\cal D}^{(j)}(g)$ is the matrix representing the $SU(2)$ element $g$ in representation $j$,
and Virasoro characters
\beq
\chi_{j^2}^{Vir}(\tilde q^2)=\frac{\tilde q^{2j^2}-\tilde q^{2(j+1)^2}}{\eta(\tilde q^2)} \ .
\eeq
Using the normalization of Ishibashi states
\beq\label{ishinorm}
\dlangle j,m,n |\Delta (\tilde q)| j',m',n'\drangle=\chi_{j^2}^{Vir}(\tilde q^2)\delta_{jj'}\delta_{mm'}\delta_{nn'}
\eeq
we obtain the boundary state in the $\widehat{SU(2)}$ basis
\beq
|N,\theta\rangle \simeq 2^{-1/4}\sum_{j=0,1/2,1,\ldots}\sum_{m,n=-j}^j
{\cal D}^{(j)}_{m,n}(e^{-2i\theta J_0^3})|j,-m,n\drangle \ .
\eeq

\subsection{Chan-Paton Factors}

With multiple branes, the above computation is only trivially modified.
The boundaries attain an extra discrete index labeling the map of the
worldsheet boundary onto a D-brane. Consequently, the boundary state
carries an index $k$, labeling the D-brane. The general construction is
reviewed more fully in the companion paper \cite{uscosmo}.

\section{Adsorption and Open String Partition Functions}\label{Adsorp}

\subsection{Boundary Deformations}

In the context of rolling tachyons, the generic boundary perturbation of interest is of the form \beq
S_{\lambda}=\int_{\partial\Sigma} ds\ \left[ \lambda_+e^{X^0(s)/\sqrt{\alpha'}} +\lambda_-e^{-X^0(s)/\sqrt{\alpha'}} \right] \eeq
where $X^0$ is the time-like target space coordinate. Classically (using the correlators of the undeformed theory), this
perturbation is marginal, that is $d_{cl}=1$. For $\lambda_{\pm}=\frac12\lambda e^{\mp X^0_0/\sqrt{\alpha'}}$, this is related to
the ``full S-brane" \cite{Gutperle:2002ai} centered at $X^0_0$, while for $\lambda_-=0$, we have the ``half S-brane"
\cite{Larsen}. The full S-brane corresponds to a process where a carefully fine-tuned initial closed string configuration time
evolves to form an unstable D-brane which then decays to a final state of closed strings
\cite{LLM,Gaiotto:2003rm,Karczmarek:2003xm,Sen:2004zm}.
The whole process is centered around the time $X^0_0$ and is time reflection invariant about it, as evident from writing the
deformation in the form \beq\label{fullS} S_{\lambda}= \lambda\int_{\partial\Sigma} ds~\cosh [(X^0(s)-X^0_0)/\sqrt{\alpha'}] \ ,
\eeq in particular the initial state of closed strings is a time reflection image of the final state. By Wick rotating $X^0=iX$,
it becomes \beq\label{fullSE} S_{\lambda}= -\lambda \int_{\partial\Sigma} ds~\cos[(X(s)-X_0)/\sqrt{\alpha'}] \ , \eeq which is a
known exactly marginal deformation \cite{Callan}. In practise, computations in the background (\ref{fullS}) are first performed in
the Euclidean signature with (\ref{fullSE}), and the results are then Wick rotated back to the Lorentzian signature. In the rest
of the present paper, we will disregard any relation to rolling tachyons, and simply consider the Euclidean theory.

One could absorb the parameter $X_0$ into the definition of the origin of time. However, for a given worldsheet with multiple
boundaries, there can be a distinct deformation for each boundary component. For example, if we consider the annulus, we will
consider a boundary deformation of the form \beql{bdydef} S_{int}= -\lambda\int_{\partial\Sigma_1} ds
\cos\left(\frac{X-X^{(1)}_0}{\sqrt{\alpha'}}\right) -\tilde\lambda\int_{\partial\Sigma_2} ds
\cos\left(\frac{X-X^{(2)}_0}{\sqrt{\alpha'}}\right) \eeq where $\partial\Sigma_j$ are the boundary components. This is essentially
a Chan-Paton structure. Indeed in the presence of multiple branes, $\lambda$ and $\tilde\lambda$ would be replaced by matrices,
and the annulus would include overall traces for each boundary component. A priori, there is no need to take the cosines to be
centred at the same point on different boundaries, and the difference cannot be absorbed to the choice of the time origin.

In the orbifold $\bR^{1,d}/\bZ_2$ the $\bZ_2$ acts by $(X^0,X^1,\ldots
,X^d)\rightarrow -(X^0,X^1,\ldots ,X^d)$ \cite{BHKN,BKLNS}. After Wick
rotation $X^0=iX$, we obtain a Euclidean orbifold $\bR^{d+1}/\bZ_2$,
where $\bZ_2$ acts by $(X,X^1,\ldots ,X^d)\rightarrow
-(X,X^1,\ldots,X^d)$. The full S-brane deformation is invariant under
the orbifold identifications, if we choose it to be centered around
$X^0=0$. In the Euclidean signature, for worldsheets with multiple
boundaries, if we allow for distinct deformations at each boundary
component $\partial\Sigma_j$, we would then need each of them to be
centered around $X=0$ (i.e., set $X^{(j)}_0=0$, but the associated
parameters $\lambda_j$ can be independent of one another. Wick rotation
back to Lorentzian signature is subtle, because of the issues with the
branching of time's arrow. This will be discussed in Ref.
\cite{uscosmo}.

\subsection{The Adsorption Method}

Let us begin with the theory at self-dual radius $R_{sd}=\sqrt{\alpha'}$.
One way to understand this case is known as
`adsorption' \cite{Affleck}. This method highlights the close relationship with the Kondo problem of
condensed matter physics.  In that situation we think of the deformation as the $U(1)$ charge current
for a free fermionic theory.


We can study this theory by replacing the boson $\hat X$ on the strip by an 'unfolded' chiral boson $\hat\phi$ on $\sigma\in
(-r,r)$, $\tau\in (0,\ell)$, where $\sigma=-r$ is identified with $\sigma=+r$ (this is a torus in the $w=i\tau+\sigma$ plane, with
modular parameter $\ell/2r$). To do so, we identify $\phi(\sigma,\tau)=X_L(i\tau+\sigma)$ for $\sigma>0$, and
$\phi(\sigma,\tau)=X_R(i\tau-\sigma)$ for $\sigma<0$. It is most natural to quantize this boson by taking $\sigma$ as the 'time'
direction; we simply have a periodic boundary condition in the $\tau$-direction, and we obtain the mode expansion \beq\hat
\phi(w)= \hat \phi_0+\frac{\pi\alpha 'w \hat p}{\ell}+i\sqrt{\frac{ \alpha'}{2}} \sum_{n\in\bZ\atop n\neq 0}\frac{1}{n} \hat
\alpha_n e^{-2\pi nw/\ell} \ . \eeq At self-dual radius, the theory reduces to a chiral $\widehat{SU(2)}$ current algebra, with
currents
\begin{eqnarray}
J^{\pm} &=& e^{\pm i\phi/\sqrt{\alpha'}} \nonumber \\
J^3 &=& i\partial \phi/\sqrt{\alpha'} \ ,
\end{eqnarray}
where $J^\pm = (J^1\pm J^2)/\sqrt{2}$. The basic strategy will be to make use of the $\widehat{SU(2)}$ current algebra relations,
\beq \mathcal{H}_0~\propto~:(J^3(\sigma))^2:~=~\frac{1}{3}:\vec{J}^2(\sigma):~=~:(J^1(\sigma ))^2: \ . \eeq For simplicity we will
focus on the situation with a single boundary deformation, taking the form \beq \lambda \int_0^\ell d\tau J^1(i\tau+0) = \ell
\lambda  J^1_0 \ . \eeq In detail, the partition function becomes: \beqn Z_\lambda  &=& Tr\
\left(e^{\frac{1}{2\pi}\int_{-r}^0d\sigma\int_0^\ell d\tau :\left(J^1(\sigma,\tau)\right)^2:}e^{i\lambda\int_0^\ell d\tau
J^1(\tau+i0)} e^{1/2\pi\int_{0}^{r}d\sigma\int_0^\ell d\tau :\left(J^1(\sigma,\tau)\right)^2:}\right) \ . \eeqn Here the boundary
deformation takes the form of an operator insertion at the fixed "time" $\sigma=0$.  Using the explicit mode expansions, we obtain
\beqn Z_\lambda &=& Tr\ \left(e^{-\frac{ r}{\ell\pi}\cdot\left(
(J^1_0)^2+2\sum_{n=1}^\infty J^1_{-n}J^1_n\right) }e^{ i \lambda J^1_0}\right)\nonumber\\
&=&\frac{1}{\eta(\tilde{q}^2)}\sum_{n\in\bZ} \left(\tilde{q}^2\right)^{n^2/4} e^{\pi i \lambda n}\ ,\ \ \ \left(\tilde{q} =
e^{-2\pi /t},\ t = l/r\right)\ . \eeqn Because we quantized using $\sigma$ as "time" we obtain an answer in the closed string
channel. We should perform a Poisson resummation to write the partition function in the open string channel, \beq Z_{\lambda} =
\sum_{m\in\bZ} \frac{q^{(m+\lambda/2)^2}}{\eta(q)}\ ,\ \ \ \left(q = e^{-\pi t},\ t=\ell/r\right)\ . \eeq

\section{Fermionic Representation}\label{fermionic}

Having studied the orbifold theory at self-dual radius, we consider now
other radii. At infinite radius, the renormalized bosonic theory may
also be represented using a `free' fermionic picture \cite{PT,KT}. We
will take the liberty in this section of discussing this construction in
detail. Although many aspects have been discussed in the literature,
certain subtle points will be needed later in the paper when we apply
the construction to orbifold theories.

The boundary interaction involves open string tachyonic vertex operators, $e^{\pm iX/\sqrt{\alpha'}}$. We should first study how
this is represented in terms of the "doubled" chiral boson living on $\sigma\in [-r,r]$.  When fermionizing we will quantize using
$\tau$ as the time direction and the mode expansion of section \ref{bdycft}, in contrast with the previous section.  With $\hat p$
conjugate to $\hat x$, this vertex operator may be written: \beqn e^{ik\hat X(\tau,0)/\sqrt{\alpha '}} &=&
e^{i2k\hat\phi(\tau)/\sqrt{\alpha '}} \\
e^{ik\hat X(\tau,r)/\sqrt{\alpha'}} &=& e^{i2k\hat\phi(\tau+r)/\sqrt{\alpha '}} e^{-i\pi\sqrt{\alpha '}k(2\hat p+k/\sqrt{\alpha
'})}\ . \eeqn In particular, when $k=\pm 1$ the vertex operator at $\sigma = r$ becomes \beq e^{\pm i\hat X(\tau,r)/\sqrt{\alpha
'}} = -e^{\pm i2\hat\phi(\tau+r)/\sqrt{\alpha '}}e^{\mp i2\pi\sqrt{\alpha '} \hat p}\ . \eeq Thus the boundary interaction is \beq
S_{int} = -\lambda\int_{\sigma=0} ds\cos(2\phi(s)/\sqrt{\alpha '})+\frac{\tilde{\lambda}}{2}\int_{\sigma=r} ds
\left(e^{i2\phi(s)/\sqrt{\alpha '}}e^{-i2\pi\sqrt{\alpha '} p}+e^{-i2\phi(s)/\sqrt{\alpha '}}e^{i2\pi\sqrt{\alpha '} p}\right).
\eeq

\subsection{Fermionic Action}

Here we want to find a fermionization appropriate for the boundary
interaction. A convenient (but not unique) way to proceed is to mix in a
second boson, $Y$. In \cite{PT} the second boson was viewed as
auxiliary. However, in the context of string theory we may use one of
the spatial directions as the second boson. It will be necessary to
introduce co-cycles in order for the algebraic properties to be
faithfully reproduced. As $X$ was related to a chiral boson $\phi$, we
may relate $Y$ to a chiral boson $\chi$.  It is possible to take
\beq
\psi_1 = e^{i(\chi-\phi)/\sqrt{\alpha '}}e^{i\pi ap}\equiv
e^{i\phi_1}e^{i\pi ap} ,\ \ \ \ \psi_2 = e^{i(\chi+\phi)/\sqrt{\alpha
'}}e^{i\pi bp}\equiv e^{i\phi_2}e^{i\pi bp},
\eeq where $a$ and $b$ are
real parameters. As before $p$ is conjugate to $x.$  We have chosen to
write the fermion cocycles in terms of the $X$ zero modes so that the
interaction is independent of the field, $Y$. We can introduce $p_\phi =
2p$, conjugate to $\phi_0.$ Similarly we may introduce $p_\chi,$
conjugate to $\chi_0.$  $\chi$ has a mode expansion similar to that of
$\phi.$

The values of $a$ and $b$ may be constrained by demanding
anticommutativity.  This leads to the condition $\frac{b+a}{2\sqrt{\alpha
'}}\in2\bZ +1.$  As in \cite{PT}, we will choose $a = 0$ and $b =
-2\sqrt{\alpha '}.$

Recall that in the doubling we have \beq \phi (\tau,r) = \phi(\tau,-r)+\pi \alpha ' p_\phi \ . \eeq Similarly, \beq \chi(\tau,r) =
\chi(\tau,-r)+\pi \alpha ' p_\chi \ . \eeq Defining parameters $\zeta_i,$ these periodicity conditions correspond to boundary
conditions on the fermions \beq \psi_i(r) = -e^{2\pi i \zeta_i}\psi_i(-r) \ . \eeq The $\zeta_i$ correspond to the fractional (in
units of $\sqrt{\alpha'}$) parts of the momenta. From the fermionization map, we may relate the $\zeta_i$ to the momenta $p_\phi$
and $p_\chi$ (mod $\mathbb{Z}$) \beqn \zeta_1 = \frac{\sqrt{\alpha'}}{2}\left(p_\chi-p_\phi\right)\ ;\ \zeta_2 =
\frac{\sqrt{\alpha '}}{2}\left(p_\chi+p_\phi\right) \ . \eeqn We will find it convenient to define $\zeta_\pm=
\frac{1}{2}\left(\zeta_1\pm\zeta_2\right)$. We then have \beq \zeta_+ = \frac{\sqrt{\alpha '}}{2}p_\chi\ ,\ \ \ \ \zeta_- =
-\frac{\sqrt{\alpha '}}{2}p_\phi \ . \eeq Given the fermionization, the interaction becomes \beq S_{int} =
-\frac{\lambda}{2}\int_\Sigma \left( \psi_1^\dagger \psi_2 e^{i\pi\sqrt{\alpha '}p_\phi} +\psi_2^\dagger\psi_1e^{-i\pi\sqrt{\alpha
'} p_\phi} \right)\delta(\sigma) +\frac{\tilde{\lambda}}{2}\int_\Sigma\left( \psi_1^\dagger\psi_2+
\psi_2^\dagger\psi_1\right)\delta(\sigma-r) \ . \eeq Unfortunately, the first term of $L_{int}$ anti-commutes with the fermion
fields.  However, it is possible to write an equivalent expression for the partition function that has a fermion number projection
operator inserted (specifically, $(1+(-1)^{F_1+F_2})/2$), while adjusting the allowed values of $\zeta_\pm$. In terms of this way
of writing the partition function, the physical states have even fermion number, and we can with impunity modify the interaction
by multiplying the first term in $L_{int}$ by $(-)^F$. These changes have no effect on the energy spectrum, but now the modified
fermionic interaction commutes with the fermionic fields. This adjustment is required for the fermionization to make sense
algebraically, and should be understood as part of the fermionization map.

Defining
\beq
\Psi= \left(
\begin{array}{r}
\psi_1 \\
\psi_2
\end{array} \right) \ ,
\eeq
we may write the full Lagrangian as\footnote{Henceforth we are setting $r=\pi$ in the range of $\sigma$.}
\beq
L = \frac{1}{2\pi}\int_{-\pi}^{\pi}d\sigma
\Psi^\dagger(\partial_\tau-\partial_\sigma +
i{\bf N}_1\delta(\sigma) -i{\bf N}_2\delta(\sigma -\pi))\Psi \ ,
\eeq
where we have defined matrices
\beq
{\bf N}_1= \pi\lambda\left[
\begin{array}{rr}
0&w \\
\overline{w}&0
\end{array} \right], \ \ \ \ {\bf N}_2 = \pi\tilde{\lambda}\left[\begin{array}{rr} 0&1\\ 1&0\end{array}\right].
\eeq
The factor $w$ is $e^{\sqrt{\alpha '}\pi i p_\phi}(-)^F.$  For fixed
values of $\zeta_\pm,$ $w$ takes the form $e^{-2\pi i \zeta_-}(-)^F.$

If we wish to study the partition function of this theory, we need to
diagonalize the Hamiltonian of the system. Following \cite{PT,KT}, we
may Fourier transform in the $\tau$ direction
\beq
\Psi(\tau,\sigma) = \int \frac{d\nu}{2\pi} e^{-i\nu \tau}\tilde{\Psi}_{\nu}(\sigma) \ .
\eeq
Inserting this into the equation of motion gives constraints on the
allowed values of $\nu $,
\beq
(-i\nu -\partial_\sigma+i{\bf N}_1\delta(\sigma)
-i{\bf N}_2\delta(\sigma-\pi))\tilde{\Psi}_\nu(\sigma)=0 \ .
\eeq
Integrating, we find
\beq\label{eq:de}
\tilde{\Psi}_{\nu}(\pi) =
e^{-2\pi i \nu}e^{-i{\bf N}_2}e^{i{\bf N}_1}\tilde{\Psi}_\nu(-\pi) \ .
\eeq
In this notation, the boundary conditions give also
\beq\label{eq:bc}
\tilde{\Psi}_{\nu}(\pi)=-e^{2\pi i(\zeta_++\sigma_3\zeta_-)}\tilde{\Psi}_{\nu}(-\pi) \ .
\eeq
To solve this equation we must demand that $\tilde{\Psi}_\nu(-\pi)$ be
proportional to an eigenvector of
\begin{eqnarray}
&& \mathcal{N}\equiv e^{-i{\bf
N}_1}e^{i{\bf N}_2}e^{2\pi i \sigma_3\zeta_-} = \\
&& \left(
\begin{array}{cc}  \bar{w}\left(\cos(\pi\lambda)\cos(\pi\tilde{\lambda})+w\sin(\pi\lambda)\sin(\pi\tilde{\lambda})\right) & iw\left(\cos(\pi\lambda)\sin(\pi\tilde{\lambda})-w\sin(\pi\lambda)\cos(\pi\tilde{\lambda})\right) \\
  i\bar{w}\left(\cos(\pi\lambda)\sin(\pi\tilde{\lambda})-\bar{w}\sin(\pi\lambda)\cos(\pi\tilde{\lambda})\right) & w\left(\cos(\pi\lambda)\cos(\pi\tilde{\lambda})+\bar{w}\sin(\pi\lambda)\sin(\pi\tilde{\lambda})\right) \\
\end{array}
\right) \ . \nonumber
\end{eqnarray}
The condition (\ref{eq:de},\ref{eq:bc}) also fixes $\nu$ up to integer shifts. The
eigenvalues of ${\cal N}$ are $e^{\pm 2\pi i \alpha}$, where
\beqn
\label{eigenval}
\sin\pi \alpha =\left(
\sin^2\left(\frac{\pi}{2}(\lambda-\tilde{\lambda})\right)\cos^2\left(\pi\zeta_-\right)+\cos^2\left(\frac{\pi}{2}(\lambda+
\tilde{\lambda})\right)\sin^2\left(\pi\zeta_-\right)\right)^{1/2}.
\eeqn
After renaming the values of $\nu$ which satisfy the equations of motion
to $\omega,$ we have the energy eigenvalues (for $\Psi$),
\beq
\nu\equiv\omega_n^{\pm} = n-\frac{1}{2} - \zeta_+\mp\alpha \ .
\eeq
We will define
\beq
\Delta_\pm = \zeta_+\pm\alpha(\zeta_-) \ .
\eeq
We write the corresponding eigenvectors as
\beq
u(\pm\alpha)\equiv\left(\begin{array}{r} u_1^{(\pm)}\\u_2^{(\pm)}\end{array}\right) \ .
\eeq

\subsection{Solutions to Fermionic EOM}

Next we present some notes on the structure of the Fock spaces. Our
reason for discussing this in so much detail is that the energy
eigenstate basis is not the natural one in the orbifold theory that we
consider later in the paper. We may write the matrix which moves us
between the $1-2$ basis and the basis in which $\mathcal{N}$ is diagonal
\beq
U = \left(\begin{array}{cc} u_1^{(+)}
&u_2^{(+)}\\u_1^{(-)}&u_2^{(-)}\end{array}\right)\equiv\left(\begin{array}{cc} \eta\cos\theta&-\xi\sin\theta\\
\xi^*\sin\theta&\eta^*\cos\theta\end{array}\right)  \ .
\eeq
When acting on the fermionic variables, we should enlarge this to
\beq
\hat{U} = \left(\begin{array}{cccc}
 u_1^{(+)} & u_2^{(+)} & 0 & 0 \\
  u_1^{(-)}& u_2^{(-)}& 0 & 0 \\
 0 & 0 & (u_1^{(+)})^*  & (u_2^{(+)})^* \\
 0 & 0 & (u_1^{(-)})^* & (u_2^{(-)})^* \end{array}\right)\ , \ {\rm when\ acting\ on}\ \left(\begin{array}{c}
  \psi_1\\
  \psi_2\\
  \psi_1^\dagger\\
  \psi_2^\dagger \end{array}\right) \ .
\eeq
With the vectors $u(\pm\alpha),$ we may write a general solution to the
equation of motion,
\beqn
\tilde{\Psi}_{\omega_n^\pm}(\sigma) &=& e^{-i\omega_n^\pm
\sigma}b_{n,\pm}\left(1|_{[-\pi..0)}
+e^{i{\bf N}_1}|_{[0..\pi)}+e^{-i{\bf N}_2}e^{i{\bf N}_1}|_{\sigma = \pi}\right)u(\pm\alpha)\nonumber \\
&\equiv&
e^{-i\omega_n^\pm \sigma}b_{n,\pm}u(\omega_n^\pm;\sigma) \ .
\eeqn
The field $\Psi(\tau,\sigma)$ then has the mode expansion
\beqn
\Psi(\tau,\sigma) &=& \sum_\pm\sum_{n\in\bZ}
e^{-i\omega_n^\pm(\tau+\sigma)}b_{n,\pm}u(\omega_n^\pm;\sigma) \ .
\eeqn
The $b_{n,\pm}$ are essentially the values of the excitations at the
$\sigma= -\pi$ boundary. Similarly, the conjugate field is written
\beq
\tilde{\Psi}^\dagger_{\omega_n^\pm}(\sigma) =
e^{+i\omega_n^\pm \sigma}\overline{b}_{n,\pm}\overline{u}(\omega_n^\pm;\sigma) \ .
\eeq
By demanding that the pieces of $\Psi$ and $\Psi^\dagger$ which vanish
as they approach the origin in the complex $z$ plane be proportional to
creation operators, we can interpret
\beqn
 b_{n,\pm} = \left\{ \begin{array}{l}
                       \ {\rm annihilation\ operator\ for}\  n>\half+\Delta_\pm \\
                       \ {\rm creation\ operator\ for}\  n <\half+\Delta_\pm
                     \end{array} \right.
\eeqn
with $\overline{b}_{n,\pm}$ as the conjugate operators with opposite action.

The normal ordered Hamiltonian is
\beqn
:H: &=& i\int d\sigma :\Psi^\dagger\partial_\tau\Psi: \nonumber \\
&=&\sum_\pm\sum_{n\in\bZ}\omega_n^\pm :\bar b_{n,\pm} b_{n,\pm}: \nonumber \\
&\equiv& \sum_\pm\left(\sum_{n\geq 1/2+\Delta_\pm}\omega_n^\pm \bar b_{n,\pm} b_{n,\pm}-
\sum_{n<1/2+\Delta_\pm}{\omega}_n^\pm b_{n,\pm}\bar b_{n,\pm} \right) \nonumber \\
&\equiv& \sum_\pm\left(\sum_{n\geq 1/2+\Delta_\pm}\omega_n^\pm \bar b_{n,\pm}
b_{n,\pm}+\sum_{n>1/2-\Delta_\pm}\bar{\omega}_n^\pm b_{1-n,\pm}\bar b_{1-n,\pm} \right) \ ,
\eeqn
with
\beqn
{\omega}_n^\pm &=& n-1/2-\Delta_\pm \ , \nonumber \\
\bar{\omega}_n^\pm &=& n-1/2+\Delta_\pm \ .
\eeqn
The vacuum of the Fock space, for given boundary conditions, then has the structure
\beq
|vac\ket = \left[\prod_{n>\frac12+\Delta_+}
|n,+\rangle
\prod_{\bar n>\frac12-\Delta_+}
\overline{|\bar n,+\rangle}
\prod_{n'>\frac12+\Delta_-}
|n',-\rangle
\prod_{\bar n'>\frac12-\Delta_-}
\overline{|\bar n',-\rangle}
\right]_{(\zeta_+,\zeta_-)} \ .
\eeq

\paragraph{Deformed Partition Function:}

With the projection onto states of even total fermion number, (before integrating) the partition function takes the form: \beqn
Z(\zeta_-,\zeta_+)&=&\frac{1}{2}\sum_{\epsilon= \pm} \prod_{\rho
=\pm}\left(q^{\frac{(\zeta_++\rho\alpha)^2}{2}-\frac{1}{24}}\prod_n\left(1+\epsilon
q^{n-1/2-\zeta_+-\rho\alpha}\right)\left(1+\epsilon q^{n-1/2 +\zeta_++\rho\alpha}\right)\right) \nonumber\\
&=&\sum_{m_-,m_+\in
\bZ} \frac{q^{(\zeta_++m_+)^2+(\alpha+m_-)^2-\frac{1}{12}}}{\prod_n (1-q^n)^2} \ .
\eeqn
and
\beq
Z = \int_0^1d\zeta_+\int_0^1
d\zeta_-Z(\zeta_+,\zeta_-) \ .
\eeq
The ranges of the $\zeta_\pm$ integrations have been carefully chosen,
given the fermion number projection, to cover the original open string
momentum space once. The $Y$ system factors out (recall $p_Y
\leftrightarrow \zeta_+$, and $\alpha$ is independent of $\zeta_+$):
\beq
Z =Z_Y\cdot\int_0^1d\zeta_-\sum_{m_-\in\bZ}\frac{q^{(\alpha +m_-)^2}}{\eta(q)} \ .
\eeq
For comparison with other computations we can write this result in the
closed string channel, via Poisson resummation,
\beqn
\label{modules}
Z &=&Z_Y\cdot\int_0^1d\zeta_-\frac{1}{\sqrt{2}}
 \sum_{\tilde{j}=0,\half,1,\ldots}\chi^{Vir}_{\tilde{j}^2}(\tilde{q}^2)\chi^{SU(2)}_{\tilde{j}}
 (e^{4\pi i\alpha J^3}) \ .
\eeqn

\subsection{Boundary States at $R=\infty$}

By observing the fermionized partition function, we can write a boundary
state which correctly reproduces the open string partition function.
Using the formula (\ref{ishinorm}) we can write the partition function
as an overlap of boundary states projected to infinite radius,
\beqn
\langle B(\tilde{\lambda};R=\sqrt{\alpha'})|P_{\infty}
\Delta(\tilde q^2)P_{\infty}|B(\lambda;R=\sqrt{\alpha'})\rangle \ ,
\eeqn
where
\beq
|B(\lambda;R=\sqrt{\alpha'})\rangle
= \sum_{j=0,1/2,1,\ldots}\sum_{m,n}\varphi^{j}
{\cal D}^{(j)}_{m,n}(e^{2\pi i\lambda J^1_0})|j,-m,n\rangle\rangle \ .
\eeq
Note that there is a possible phase $\varphi$ here that is undetermined
by the annulus computation; we will retain it for now. The projection to
infinite radius is defined as \cite{GR}
\beqn
P_{\infty}|B(\lambda;R=\sqrt{\alpha'})\rangle=
\int_0^1d\zeta_-\sum_{j=0,1/2,1,\ldots}\varphi^{j}
\sum_{m,n} {\cal D}^{(j)}_{m,n}
(e^{+2\pi i\zeta_- J^3_0}
e^{2\pi i\lambda J^1_0}e^{+2\pi i\zeta_- J^3_0})|j,-m,n\rangle\rangle \ .
\eeqn
Since $P_{\infty}$ is a projection operator, one may show that this
boundary state is consistent with the form of
$Z_{\lambda,\tilde{\lambda}}$. We may also show that this is equivalent
to the standard expression \cite{Callan}
\beq
P_{\infty}|B(\lambda;R=\sqrt{\alpha'})\rangle =
\sum_{j=0,1/2,1,\ldots}\varphi^{j}\sum_{m} {\cal D}^{(j)}_{-m,m}(e^{2\pi i\lambda J^1_0})|j,m,m\rangle\rangle \ ,
\eeq
where it is manifest that the winding modes have been projected out by
the infinite radius limit.

Let us also check various limits of
$\lambda$. First, take $\lambda =0$, which should give us the Neumann
state:
\beq
P_{\infty}|B(0;R=\sqrt{\alpha'})\rangle =
\sum_{j=0,1/2,1,\ldots}\varphi^{j} |j,0,0\rangle\rangle \ ,
\eeq
where we have used\footnote{In the notation used here, the general formula is
\[
{\cal D}^{(j)}_{mn}\begin{pmatrix}a&b\\ c&d\end{pmatrix} =
\sum_k \frac{[(j+m)!(j-m)!(j+n)!(j-n)!]^{1/2}}{k!(j-m-k)!(j+n-k)!(m-n+k)!}
a^{j+n-k}b^{m-n+k}c^kd^{j-m-k}.\]}
${\cal D}^{(j)}_{m,n}(\mathbb{I})=\delta_{m,n}$.
This is the expected result.

For $\lambda=1/2$:
\beq
P_{\infty}|B(1/2;R=\sqrt{\alpha'})\rangle =
\sum_{j=0,1/2,1,\ldots}\sum_{m}\varphi^{j} {\cal D}^{(j)}_{-m,m}(e^{i\pi J^1_0})|j,m,m\rangle\rangle
\eeq
and one finds ${\cal D}^{(j)}_{-m,m}(e^{i\pi J^1_0})=e^{i\pi j}$. Thus we have
\beq
P_{\infty}|B(1/2;R=\sqrt{\alpha'})\rangle =
\sum_{j=0,1/2,1,\ldots}\varphi^{j}e^{i\pi j}\sum_{m} |j,m,m\rangle\rangle \ .
\eeq
By rewriting this expression in oscillator variables, it becomes clear
that it corresponds to an {\it array} of point-like D-branes separated
by a distance $2\pi\sqrt{\alpha'}$; that is, it is proportional to
\cite{Sen} \[\sum_{s\in\bZ} \delta(x-(\pi/2+2\pi s)\sqrt{\alpha'})\ ,\] if
we set the phase $\varphi$ to one. In fact, $\varphi$ simply corresponds
to the freedom to translate the array of branes.

\subsection{Projection to generic radii}\label{sec:genrad}

The fermionic calculation considered above is, strictly speaking, valid
at infinite radius. The bosonic calculation is on the other hand valid
at self-dual radius. With some care, we can in fact do the fermionic
calculation at any (rational) radius. To our knowledge, this has not
been described before from the open string point of view. In the
boundary state formalism, there is a proposal \cite{GR,GRW}, given by
introducing suitable projection operators. By carefully considering the
open string calculation, we will be able to explicitly display the
meaning of these projection operators, and the limits of applicability
of the fermionic picture.

To see that there is a potential problem in the fermionic theory,
consider the fermion boundary conditions. Recall that $\zeta_-$ is the
fractional part of what was open string momentum in the bosonic picture.
At finite radius, the open string momentum is quantized in units of
$1/R$, and thus one might expect that one could obtain the annulus
amplitude by restricting the values of $\zeta_-$ appropriately
\cite{PT}. However, this only has limited applicability, to the case of
integer radius (in units of $\sqrt{\alpha'}$). One can easily show that
this procedure fails for any other radius.

In the fermionic picture this can be seen from the boundary conditions:
if the radius is not an integer, then $\zeta_-=1$ is not equivalent to
$\zeta_-=0$, and this means that the fermionization does not make sense.
This has roots in the fact that in the bosonic picture, the boundary
operator $e^{iX/\sqrt{\alpha'}}$ is not single-valued at generic radius,
and thus the deformed theory does not exist. It is easy to repair this
however, at least at rational radius, as we will now show.

Indeed, we may think of the finite radius theory as a shift orbifold,
and define the theory by introducing a projection operator into
open-string correlators of the form
\beq
P_{(R)}=\sum_{m\in\bZ} e^{2\pi i\hat P m R} \ ,
\eeq
where $\hat P$ is the momentum operator. In the undeformed theory, this implements the projection to finite radius correctly. In the deformed theory, the boundary operator
\beq
S_{\lambda}=\frac12\int_{\partial\Sigma} ds\ \left[ \lambda e^{iX(s)/\sqrt{\alpha'}}+ h.c. \right]
\eeq
undergoes a transformation
\beq
S_{\lambda}\to\frac12\int_{\partial\Sigma} ds\ \left[ \lambda e^{2\pi iR/\sqrt{\alpha'}} e^{iX(s)/\sqrt{\alpha'}}+ h.c. \right]
\eeq
under the shift $X\to X+2\pi R$. Thus, if we insert $P_{(R)}$ into the
path integral, it does not commute with the action precisely, and so the
theory is not well-defined. It is easy to see how to repair this
however; essentially, in the deformed BCFT, we must include a
non-trivial action in Chan-Paton corresponding to the shift. We may
define a new theory by simply averaging the infinite radius theory over
the values of $\lambda$ in the image of all possible shifts. If we write
$R=(M/N)\sqrt{\alpha'}\equiv r\sqrt{\alpha'}$, then we would have for
the annulus amplitude
\beq\label{eq:finrad}
{\cal A}_{R;\lambda,\tilde\lambda}\equiv \frac{1}{N^2}\sum_{n=0}^{N-1}\sum_{\tilde n=0}^{N-1}
Tr (P_{(R)} q^{L_o-1/24} e^{S_{\lambda e^{2\pi inr}}+S_{\tilde\lambda e^{2\pi i\tilde n r}}}) \ .
\eeq
This theory has the interpretation of the Chan-Paton space for each
boundary being extended to be $N$-component, each with a complex
deformation parameter $\lambda e^{2\pi i nr}$, for $n=0,1,...,N-1$. Note
that at integer radius ($N=1$), this modification has no effect.

We emphasize that the expression (\ref{eq:finrad}) is an infinite radius
calculation, expressed in the bosonic language. It is natural, because
of the insertion, to evaluate it in momentum space. In this case, eq.
(\ref{eq:finrad}) differs from previous computations in two ways: first,
there is a momentum dependent phase factor, and secondly, for any given
$m,n,\tilde n$, there are effectively complex values of the boundary
deformation parameters. Since we are at infinite radius, we may
fermionize. In so doing, we find a generalization of the previous
result: the important parameter $\alpha$ now takes the form
\beq\label{eq:newalpha}
\sin^2\pi\alpha = \sin^2\frac{\pi}{2}(|\lambda|-|\tilde\lambda|)
+\sin^2\pi\zeta_-\cos\pi |\lambda|\cos\pi |\tilde\lambda|
+\sin^2\pi(n-\tilde n)r\sin\pi |\lambda|\sin\pi |\tilde\lambda|
\eeq
depending in general on $|\lambda|,n,|\tilde\lambda|,\tilde n,\zeta_-$.

We recall that the open string momentum was split into an integer $\ell$
and $\zeta_-\in [0,1)$. Thus, we obtain
\beq
{\cal A}_{R;\lambda,\tilde\lambda}=\frac{1}{N^2\eta(q)}
\sum_{m,\ell\in\bZ}\sum_{n=0}^{N-1}\sum_{\tilde n=0}^{N-1}
\int_0^1 d\zeta_-\ e^{2\pi i\zeta_- mr}e^{2\pi i\ell mr}  q^{(\ell+\alpha)^2}
\eeq
with $\alpha$ now given by eq. (\ref{eq:newalpha}), which is the
generalization of previous results to complex couplings. Note that it is
encoding the fact that $|\lambda|$ and $|\tilde\lambda|$ have been
renormalized (in the fashion given by Ref. \cite{Callan}) but the phases
$n$, $\tilde n$ are essentially not renormalized.\footnote{The reason
for this dichotomy is that the renormalization comes from the collision
of $J_+$ and $J_-$ insertions, each of which is accompanied by
$|\lambda|$, but opposite phases. Thus the powers of $|\lambda|$ build
up, but the phases tend to cancel and do not renormalize.} Thus the
result is a function of $|\lambda|,n;|\tilde\lambda|,\tilde n$ but {\em
not} just of $\lambda,\tilde\lambda$.

It is important to realize that what we have done here is to resolve the
classical problem of finite radius. In going to the fermionic
representation however, there is a potential quantum problem as well --
that is, the fermionic measure may not be well-defined in the presence
of the projection operator. Indeed, the translation operator corresponds
precisely to a chiral $\bZ_{2N}$ transformation on the fermions, and
thus the measure is not invariant, transforming by a $\bZ_N$ phase. This
may be repaired by introducing an array of Wilson lines, as
\beq\label{eq:ARLL}
{\cal A}^{fermionic}_{R;\lambda,\tilde\lambda}=\frac{1}{N^2\eta(q)}
\sum_{m,\ell\in\bZ}\sum_{n=0}^{N-1}\sum_{\tilde n=0}^{N-1}\int_0^1 d\zeta_-\ e^{2\pi i\zeta_- mr}
e^{2\pi i\ell mr}  q^{(\ell+\alpha)^2} \sum_{k=0}^{N-1}e^{-2\pi i mk/N} \ .
\eeq
The idea is that the action of the projection operator on the fermionic
measure leads to a $\bZ_N$ phase, which can be absorbed by a shift to
another Wilson line. Thus, we obtain a well-defined fermionic theory by
summing over such Wilson lines.  It is important to note here that what
is being said is that consistent fermionic theories can be defined only
if we include the array of Wilson lines. The bosonic theory exists in
the absence of the Wilson line array, but we have no way to compute in
the deformed theory, away from self-dual radius. Note further that the
Wilson line array appears only for fractional radii (i.e., $N\neq 1$);
the integer radius cases work just fine without it.

Now what the array of Wilson lines does is force $m=Ns$, for $s\in\bZ$.
We then obtain
\beqn
{\cal A}^{fermionic}_{R;\lambda,\tilde\lambda}&=&\frac{1}{N^2\eta(q)}\sum_{s,\ell\in\bZ}\sum_{n=0}^{N-1}
\sum_{\tilde n=0}^{N-1}\int_0^1 d\zeta_-\ e^{2\pi i\zeta_- Ms}  q^{(\ell+\alpha)^2} \nonumber \\
&=&\frac{1}{MN^2\eta(q)}\sum_{\ell\in\bZ}\sum_{n=0}^{N-1}\sum_{\tilde n=0}^{N-1}\int_0^1 d\zeta_-\
\sum_{k'\in\bZ}\delta(\zeta_- -k'/M)  q^{(\ell+\alpha)^2} \nonumber \\
&=&\frac{1}{MN^2\eta(q)}\sum_{\ell\in\bZ}\sum_{n,\tilde n=0}^{N-1}
\sum_{k'=0}^{M-1}  q^{(\ell+\alpha(\zeta_-=k'/M))^2} \ .\label{finalres}
\eeqn
For later use, we note that this can be manipulated into the form
\beq
\frac{1}{MN\eta(q)}\sum_{\ell\in\bZ}\sum_{k=0}^{N-1} \sum_{k'=0}^{M-1}  q^{(\ell+\beta(\zeta_-=k'/M))^2}
\eeq
where
\beq
\cos 2\pi\beta=\cos\pi|\lambda|\cos\pi|\tilde\lambda| \cos 2\pi\zeta_- +\sin\pi|\lambda|\sin\pi|\tilde\lambda| \cos 2\pi k/N
\eeq
 This formula is a direct consequence of (\ref{eq:newalpha}).

Let us consider a few special cases. First, the NN amplitude should be
recovered by taking $\lambda,\tilde\lambda\to 0$. In this case,
$\alpha=\zeta_-$, independent of $n,\tilde n$, and so we find
\beq
\frac{1}{M\eta(q)}\sum_{\ell\in\bZ} \sum_{k'=0}^{M-1}  q^{(\ell+k'/M)^2}
=\frac{1}{M\eta(q)}\sum_{\ell\in\bZ}  q^{\ell^2/M^2}
\eeq
which may be recognized as the NN amplitude at radius $R$, with the
array of Wilson lines. Note also that
\beq
{\cal A}_{R;1,0}= \frac{1}{M\eta(q)}\sum_{\ell\in\bZ} q^{(\ell/M+1/2)^2} \ .
\eeq

Next let us consider the ND case, obtained by $\lambda=1/2$,
$\tilde\lambda=0$. Here, we obtain $\alpha=1/4$, independent of
$n,\tilde n$, as well as $\zeta_-$ and thus
\beq
{\cal A}_{R;1/2,0}=\frac{1}{\eta(q)}\sum_{l\in\bZ} q^{(\ell+1/4)^2} \ .
\eeq

Finally consider the DD case, obtained via $\lambda=\tilde\lambda=1/2$,
whence $\alpha=(n-\tilde n)r$. Thus, we find
\beq
{\cal A}_{R;1/2,1/2}=\frac{1}{N^2\eta(q)}\sum_{\ell\in\bZ}\sum_{n,\tilde n=0}^{N-1}
q^{(\ell+(n-\tilde n)r)^2} \ .
\eeq
It is possible to show that the sum over $n,\tilde n$ can be written as
\beq
{\cal A}_{R;1/2,1/2}=\frac{1}{N\eta(q)}\sum_{\ell\in\bZ}\sum_{k=-j}^{j} q^{(N\ell+kM)^2/N^2}
\eeq
where $j=[[ N/2]]$. This is
\beq
{\cal A}_{R;1/2,1/2}=\frac{1}{N\eta(q)}\sum_{\ell\in\bZ} q^{\ell^2/N^2}
\eeq
which is the correct result for an array of $M$ D-branes, located at
integer multiples of $\sqrt{\alpha'}/N$. This is the expected result
\cite{Sen}, with branes located at extrema of the boundary potential. It
is clear that this array of D-branes is T-dual to the array of Wilson
lines at $\lambda=\tilde\lambda=0$. For generic values of
$\lambda,\tilde\lambda$, we interpolate smoothly between the two arrays,
again as should be expected. The absence of the fermionic anomaly
mentioned above corresponds to the recovery of T-duality in the annulus
amplitude.

\subsubsection{Boundary States at Radius $r$}

These finite radius annulus amplitudes may be transformed into the
closed string channel. The result is consistent with boundary states
given (from closed channel reasoning) by Gaberdiel and Recknagel
\cite{GR}. They are written via projection operators acting on the
self-dual radius result (here we have made a requisite translation into
our conventions)
\beqn
|B(\lambda; R)\rangle &=& P^+_MP^-_N\sum_{j\in 0,1/2,..}\sum_{m,\tilde{m}=-j}^{j}\varphi^{j}{\cal
D}^j_{m\tilde{m}}\left(e^{2\pi i \lambda
J^1}\right)|j,-m,\tilde{m}\drangle \\
&\equiv& \frac{1}{\sqrt{MN}}\sum_{\ell= 0}^{M-1}\sum_{k=0}^{N-1}\sum_{j\in 0,1/2,..}\sum_{m,\tilde{m}=-j}^{j}\varphi^{j}{\cal D}
^j_{m\tilde{m}}\left(e^{2\pi i(\frac{\ell}{M}+\frac{k}{N})J^3}e^{2\pi i \lambda
J^1}e^{2\pi i(\frac{\ell}{M}-\frac{k}{N})J^3}\right)|j,-m,\tilde{m}\drangle.\nonumber
\eeqn

Using this boundary state, one obtains the following for the open string
partition function,
\beq
\frac{1}{MN}\sum_{l=0}^{M-1}\sum_{k=0}^{N-1}\sum_{n\in\bZ}\frac{q^{(n+\beta(\lambda,\tilde{\lambda};\ell,k))^2}}{\eta(q)}.\eeq
Here $\beta$ satisfies
\beq
2\cos(2\pi\beta(\lambda,\tilde{\lambda};\ell,k))\equiv Tr_{1/2}\left(e^{-2\pi i
\tilde{\lambda}J^1}e^{2\pi i(\frac{\ell}{M}+\frac{k}{N})J^3}e^{2\pi i\lambda J^1}e^{2\pi i(\frac{\ell}{M}-\frac{k}{N})J^3}\right).
\eeq
One may show that the set of values of $\beta$ is equivalent to the set
of $\alpha$'s in eq. (\ref{finalres}). This result is equivalent to the
open string result, as long as the array of Wilson lines is included.
Note that this is important, because the boundary states proposed
transform in a simple way under T-duality. Without the Wilson lines,
T-duality is not implemented by flowing from $\lambda=0$ to
$\lambda=1/2$.

\section{The $\bZ_2$ Orbifold}\label{orbifoldsec}

We will now consider the orbifold action in the BCFT (\ref{eq:bdydef}).
For the annulus computation, this is implemented by including the
orbifold projection operator $\frac12(1+g)$ in the path integral. The
first term (proportional to $1$) is equivalent, up to the factor of two,
to the results given above. On a self-dual circle we may derive the
effect of inserting $g$ through adsorption methods (see Ref.
\cite{AfflOsh} for the construction in the undeformed theory).  Because
the states are classified by current algebra and we know the orbifold
action on the algebra, it is straightforward to compute the $g$-inserted
trace at self-dual radius. As in the previous sections, to discuss more
general radii we must fermionize the theory. When fermionizing we will
consider two possible orbifold actions: either $g$ acts only the $X$
field (``$c=1$ orbifold") or it acts on both $X$ and $Y$ (``$c=2$
orbifold"). In either case, we should be able to disentangle the
($g$-inserted) partition functions of the $X$ and $Y$ systems. The
consistency of these three routes is strong evidence that we have
correctly orbifolded the deformed boson.

In the orbifold theory, there are a variety of open string annulus
calculations that we can do, depending on the details of Chan-Paton
factors, as we have discussed briefly above. We will consider these
details in Ref. \cite{uscosmo} and here simply concentrate on the
calculation of the annulus diagram with orbifold insertion. This will be
the basic building block needed to construct fractional boundary states.

\subsection{Self-dual Radius and Adsorption}\label{orbAdsorp}

The orbifold is obtained by
defining $\bZ_2$ as $X\mapsto -X$, or in terms of the current,
\beq
   J^1 \rightarrow J^1 \ ; \ J^2 \rightarrow -J^2 \ ; \ J^3 \rightarrow -J^3 \ .
\eeq

At self-dual radius, it is convenient to organize the calculation entirely in terms of the $\widehat{su(2)}_1$ modules.
Since the deformation is in the direction
of $J^1$ rather than
$J^3$, it is more convenient to work in the $su(2)$ basis
where $J^1_0$ is diagonal, as we described in Section \ref{Adsorp}. The orbifold action then switches the sign of the ladder operators.
The $g$-inserted partition function is (see Appendix A for a detailed analysis)
\begin{eqnarray}
\Tr gq^{L_0-1/24}=\frac{1}{\eta(q)}\sum_{n\in\bZ}(-1)^{n}q^{(n+\lambda/2)^2} \ .
\end{eqnarray}
For other radii, we now turn to the fermionic description.

\subsection{Fermionic Description of the Orbifold}

Let's discuss the $\bZ_2$ orbifold of the deformed theory.  After
detailed computations, the two orbifold actions may be determined
consistently to be as given in the following table.
\[\begin{tabular}{|c|c|}
  \hline
  $c=2$ Orb & $c=1$ Orb \\
  \hline
  $X\rightarrow -X$ & $X\rightarrow -X$ \\
  $Y\rightarrow -Y$ & $Y\rightarrow Y$ \\
  $\psi_1\rightarrow\psi_1^\dagger$&
  $\psi_1\rightarrow i\psi_2e^{2\pi i\zeta_-}$\\
  $\psi_2\rightarrow -\psi_2^\dagger$&
  $\psi_2\rightarrow i\psi_1e^{2\pi i\zeta_-}$\\
  $\psi_1^\dagger\rightarrow\psi_1 $&
  $\psi_1^\dagger\rightarrow -i\psi_2^\dagger e^{-2\pi i\zeta_-}$\\
  $\psi_2^\dagger\rightarrow -\psi_2$&
  $\psi_2^\dagger\rightarrow -i\psi_1^\dagger e^{-2\pi i \zeta_-}$\\
  \hline
\end{tabular}\]

The phases that appear in the table may look strange; in particular, it
might appear that $g_1$ does not square to one. However, the table
refers to the action on single fermions, which do not by themselves
create physical states (recall the fermion number projection). The
phases in the table are fixed by the requirement that the $SU(2)_X$
current algebra as well as other fermion bilinears transform in a
sensible way under the orbifold actions. In particular, $g$ does square
to one on all physical states, in each case.

In order to proceed we need to rephrase the orbifold as an action in the
$\pm$ basis rather than the $1-2$ basis.  This is accomplished by making
a similarity transformation on the orbifold generator, $\hat{U}g
\hat{U}^\dagger$. The action of $g_1$ on the $\psi-\psi^\dagger$ and
$\pm$ labels of the field is given by:
\beq
 \hat Ug_1\hat U^{-1}=
\begin{pmatrix}
iG_1 &0\cr 0& -iG_1^*
\end{pmatrix}
\eeq
where
\beq
G_1=e^{2\pi i\zeta_-}UgU^{-1} \ ,
\eeq with $g$ from the table,
\beq
g=\begin{pmatrix} 0&1\cr 1&0\end{pmatrix} \ ,
\eeq
and $U$ was given in a preceding section.
The action of $g_2$ on the $\psi-\psi^\dagger$ and $\pm$
labels of the field is given by:
\beq
\hat Ug_2\hat U^{-1}=
\begin{pmatrix}
0&G_2 \cr  G_2^* &0
\end{pmatrix}
\eeq
where
\beq
G_2=UgU^{-1}
\eeq and from the table,
\beq g=\begin{pmatrix} 1&0\cr
0&-1\end{pmatrix} \ .
\eeq

Having determined how the orbifold acts on the fields of the theory, it
remains to determine the action on the ground state. Fortunately, this
is facilitated by the fact that we knew how the orbifold acted upon the
momenta of the bosonic theory. Consistency demands that $g$ acts on the
$\zeta_-$ and $\zeta_+$ as it did on the $X$ and $Y$ momenta,
respectively.  The orbifold action on the interacting ground state is:
\beqn
g_1\cdot
\left[
|n,+\rangle \otimes
\overline{|\bar n,+\rangle}\otimes
|n',-\rangle\otimes
\overline{|\bar n',-\rangle}
\right]_{(\zeta_+,\zeta_-)}=\left[
|n,+\rangle \otimes
\overline{|\bar n,+\rangle}\otimes
|n',-\rangle\otimes
\overline{|\bar n',-\rangle}
\right]_{(\zeta_+,-\zeta_-)}\nonumber\\
g_2\cdot
\left[
|n,+\rangle \otimes
\overline{|\bar n,+\rangle}\otimes
|n',-\rangle\otimes
\overline{|\bar n',-\rangle}
\right]_{(\zeta_+,\zeta_-)}=\left[
|n,+\rangle \otimes
\overline{|\bar n,+\rangle}\otimes
|n',-\rangle\otimes
\overline{|\bar n',-\rangle}
\right]_{(-\zeta_+,-\zeta_-)}\nonumber
\eeqn
We see that $g_1$ effectively flips the sign of $\zeta_-$ and $g_2$
flips the sign of both $\zeta_+$ and $\zeta_-$ inside the vacuum states.

\subsubsection{Orbifold-inserted Traces}

Because the orbifold action mixes all of the oscillators, at each value
of $n$ there is the possibility for many terms to contribute. The
orbifold actions on the vacuum states limit the values of $\zeta_{\pm}$
which contribute to the $g$-inserted partition functions--that is, there
will be $\delta$-functions which restrict to the fixed points of the
orbifold action. As before, we use the same fermion insertion
$(1+(-1)^{F_T})/2$ and integration measure
$\int_0^1d\zeta_+\int_0^1d\zeta_-$.

\paragraph{$c=1$ Orbifold:}

Since $\zeta_+\mapsto \zeta_+$ and $\zeta_-\mapsto-\zeta_-$ under the
orbifold action, there is a fixed line at $\zeta_-=0$. Here,
$\alpha=(\lambda-\tilde\lambda)/2$ (mod 1), and
\beq
U=\begin{pmatrix}\frac{1}{\sqrt{2}}&-\frac{1}{\sqrt{2}}\cr
\frac{1}{\sqrt{2}}&\frac{1}{\sqrt{2}}\end{pmatrix}
\eeq
and thus the orbifold action on single fermion states is determined by
\beq
\hat{U}g_1\hat{U}^\dagger=
\begin{pmatrix} iG&0\cr 0& -i G^*\end{pmatrix}=
\left(
\begin{array}{cccc}
-ie^{2\pi i\zeta_-}&0&0&0\\
0&ie^{2\pi i\zeta_-}&0&0\\
0&0&ie^{-2\pi i\zeta_-}&0\\
0&0&0&-ie^{-2\pi i\zeta_-}\end{array}
\right) \ .
\eeq
Note that the factors of $i$ that appear here are required by
consistency (although they did not appear "geometrically"). As we
discussed above, it looks as if $g_1^2=-1$; however, this is acting on
single-fermion states--since there are no such physical states, we can
allow such an action. It must act as $g_1^2=+1$ on all double fermion
states however. We can see that we need the factors of $i$ by examining
operators like $\psi_+\psi_-$, which consists of $Y$ only, and thus
should be orbifold invariant.

There is also a fixed line at $\zeta_-=1/2$, because $\zeta$'s are
defined mod 1. Here, $\alpha=(\lambda+\tilde\lambda-1)/2$ (mod 1) and
$U$ is the same as above.

The $g_1$ trace takes the form $\int_0^1 d\zeta_+\int_0^1d\zeta_-
\left[\delta(\zeta_-)+\delta(\zeta_--1/2)\right]$ times
\[
q^{\zeta_+^2+\alpha^2-1/12}\frac12\sum_\pm
\prod_{n=1}^\infty (1\mp ie^{2\pi i\zeta_-}q^{\omega_n^+}) (1\pm ie^{2\pi i\zeta_-}q^{\omega_n^-}) (1\pm
ie^{-2\pi i\zeta_-}q^{\bar\omega_n^+}) (1\mp ie^{-2\pi i\zeta_-}q^{\bar\omega_n^-}) \ .
\]
The sum $\sum_\pm$ is the fermion number projection. This result can be
written in terms of a product of $\theta$-functions; using the sum
representation for the $\theta$-functions, we then find
\[ \frac{q^{\zeta_+^2+\alpha^2}}{\eta^2(q)}\sum_{n,m\in\bZ} q^{n^2/2}q^{m^2/2}
(ie^{2\pi i\zeta_-}q^{-\zeta_+-\alpha})^n (ie^{2\pi i\zeta_-}q^{-\zeta_++\alpha})^m\left[ \frac{(-1)^n+(-1)^m}{2}\right]\]
Defining $m_\pm=(n\pm m)/2$, we find
\beqn
Z^{g_1}_{\lambda,\tilde{\lambda}}&=&\int d\zeta_+\sum_{\zeta_-=0,1/2}\frac{q^{\zeta_+^2+\alpha^2}}{\eta^2(q)}\sum_{m_+,m_-\in\bZ}
q^{m_+^2}q^{m_-^2} e^{4\pi i\zeta_-m_+} q^{-2\zeta_+m_+} q^{-2\alpha m_-} (-1)^{m_-}\nonumber\\
&=& \left(\frac{1}{\eta(q)}\int_0^1 d\zeta_+\sum_{m_+\in\bZ} q^{(m_++\zeta_+)^2}\right)\cdot\left(\sum_{\zeta_-=0,1/2}
\frac{1}{\eta(q)}\sum_{m_-\in\bZ}q^{(m_-+\alpha)^2} (-1)^{m_-}\right) \ .
\eeqn
This is to be summed over the two values of $\zeta_-$; for both of those
values, the $\zeta_-$-dependent term in the $m_+$ sum equals unity, and
we have dropped it.  So, we see that the $g_1$-inserted partion function
decouples nicely into $X$ and $Y$. The integral over $\zeta_+$ combines
with the $m_+$ sum to give the partition function of the free boson $Y$
while the $m_-$ sum is the  twisted partition function of $X$.

\paragraph{$c=2$ Orbifold:}
Because both $\zeta_\pm$ are set to zero modulo periodicities in the
$c=2$ orbifold, we only get contributions from the points $(0,0)$ and
$(1/2,1/2).$  The action of the orbifold on the fermion fields is given
by:
\beq
\hat{U}g_2\hat{U}^\dagger=\left(\begin{array}{cccc}0&0&0&1\\0&0&1&0\\0&1&0&0\\1&0&0&0\end{array}\right) \ .
\eeq
The diagonal combinations of states are
$$
|vac\rangle \ ,\ \psi^\dagger_-\psi_+|vac\rangle \ , \ \psi^\dagger_+\psi_-|vac\rangle \ , \
 \psi^\dagger_-\psi_+\psi^\dagger_+\psi_-|vac\rangle \ .
$$
For the two points, we find
$$
{\rm at}\ \zeta_\mp
= 0: \  \alpha(0) = \frac{\lambda-\tilde{\lambda}}{2} \ ,
\  \ {\rm Casimir\ energy} =
\left(\frac{\lambda-\tilde{\lambda}}{2}\right)^2-\frac{1}{12}
$$
$$
{\rm at}\ \zeta_\mp
= 1/2: \  \alpha(1/2) = \frac{\lambda+\tilde{\lambda}-1}{2} \ ,
\  \ {\rm Casimir\ energy} =
\left(\frac{\lambda+\tilde{\lambda}-1}{2}\right)^2-\frac{1}{12} \ .
$$
%
The trace, with $g_2$ inserted, becomes:
\beqn Z^{g_2}_{\lambda,\tilde{\lambda}} &=&
    \sum_{(\zeta_-,\zeta_+)=(0,0),(1/2,1/2)}q^{\alpha^2-1/12}\prod_n\left(1-(q^2)^{
    n-1/2+\alpha}\right)\left(1-(q^2)^{n-1/2-\alpha}\right)\nonumber\\
&=&  \left(\frac{q^{-1/24}}{\prod_n\left(1+q^{n}\right)}\right) \cdot\left(
   \sum_{\zeta_-=0,1/2}\frac{1}{\eta(q)} \sum_{n\in\bZ}(-)^n q^{(n+\alpha)^2}\right) \ .
\eeqn

We have split the trace into separate contributions from
$(\zeta_-,\zeta_+)\ =(0,0)$ and $(1/2,1/2).$ The minus sign in the
factors is due to the anti-commutivity of the fermion fields. Note that
the powers of $q$ here are all independent of $\zeta_+$, an important
feature. The insertion of $g_2$ restricts the trace to be over states of
even fermion number, implying that the total fermion number projection
operator acts as the identity in the presence of $g_2$. Again, the
$g$-inserted partition function has factorized into contributions from
$X$ and $Y$.

\subsection{Summary}
At the self-dual radius, one finds
\beq
Z^g_{\lambda,\tilde{\lambda}}
= \frac{1}{\eta(q)}\sum_{n\in\bZ}(-)^n q^{(n+(\lambda-\tilde\lambda)/2)^2}
\eeq
using adsorption methods. For infinite radius, we may decouple the
$Y$-system from the result of fermionization to obtain
\beq
Z_1 = \frac{1}{\eta(q)}\int_0^1 d\zeta_-\sum_{m\in\bZ} q^{(m+\alpha(\zeta_-))^2}
\eeq
and
\beq
 Z_{g}=\frac{1}{\eta(q)}\sum_{\zeta_-=0,1/2}\sum_{m\in\bZ}(-1)^{m}q^{(m+\alpha(\zeta_-))^2} \ .
\eeq
Here,
\beqn
\sin\pi \alpha =\left(
\sin^2\left(\frac{\pi}{2}(\lambda-\tilde{\lambda})\right)\cos^2\left(\pi\zeta_-\right)+
\cos^2\left(\frac{\pi}{2}(\lambda+
\tilde{\lambda})\right)\sin^2\left(\pi\zeta_-\right)\right)^{1/2}.
\eeqn

Notice that the $\zeta_-=0$ term reproduces the self-dual radius result.
This is correct, since at self-dual radius half-integer momentum
($\zeta_-=1/2$) is not present. It is a confirmation of our methods that
the contributions at $\zeta_-=0$ to $Z^g_{\lambda,\tilde\lambda}$
obtained from fermionization match the result of the self-dual radius
adsorption methods.


Also recall that for $Z_1$, we can write the result in terms of Virasoro
and $SU(2)$ characters of the closed string channel
\beqn
Z_1&=&\frac{1}{\sqrt{2}}\frac{1}{\eta(\tilde q^2)}\int_0^1 d\zeta_-
\sum_{m\in\bZ}(\tilde q^2)^{m^2/4}e^{2\pi i\alpha(\zeta_-)m} \nonumber \\
&=&\frac{1}{\sqrt{2}}\sum_{j=0,1/2,1,\ldots}\chi^{Vir}_{j^2}(\tilde q^2)\int_0^1 d\zeta_-
\chi_j^{SU(2)}(e^{4\pi i\alpha(\zeta_-)J^3}) \ .
 \eeqn
The corresponding result for $Z_g$ follows from Poisson resummation
\beq
Z_g=\frac{1}{\sqrt{2}}\frac{1}{\eta(\tilde q^2)}\sum_{\zeta_-=0,1/2}
\sum_{m\in\bZ}(\tilde q^2)^{(m-1/2)^2/4}e^{2\pi i\alpha(\zeta_-)(m-1/2)} \ .
\eeq
Other radii may be obtained by suitable projection operators, as
discussed in Section \ref{sec:genrad}. A treatment of all possible
boundary states of the orbifold theory appears in Ref. \cite{uscosmo}.

\section{Conclusions}

In this paper, we have considered exactly marginal boundary deformations
of the $c=1$ theory through one-loop computations directly in the open
string channel. The partition function at generic radius was constructed
through a fermionization technique and shown to coincide with expected
boundary states. In doing so, it was necessary to carefully define the
fermionic theory so as to be consistent with the projection to finite
radius. We have also carefully constructed the orbifold $\bR^d/\bZ_2$
theory in the fermionic parameterization and computed the corresponding
twisted partition function. Further results and applications of our
results will appear in Ref. \cite{uscosmo}.

\bigskip

\noindent
{\Large \bf Acknowledgments}

\medskip

\noindent EKV thanks UCLAs IPAM and the organizers of the Conformal Field Theory 2nd Reunion Conference, and the organizers of the
Workshop on Gravitational Aspects of String Theory at Fields Institute for hospitality while this work was in progress.  SK would
like to thank Matthias Gaberdiel, Andreas Recknagel, and Masaki Oshikawa for discussions, and also the organizers of the July 2005
London Mathematical Society Durham Symposium on "Geometry, Conformal Field Theory and String Theory."  EKV was in part supported
by the Academy of Finland. SK was in part supported by a JSPS fellowship. RGL and SN have support from the US Department of Energy
under contract DE-FG02-91ER40709.

\bigskip

\noindent
{\Large \bf Appendix A: }

\bigskip

For convenience, we present more details of the SU(2) module calculations of Sections \ref{Adsorp} and \ref{orbAdsorp} here.
The multiplicities of the $j=0$ SU(2) module can be found in \cite{BYB}, Fig. 15.1. By direct computation, we obtain
\begin{eqnarray}
tr\ q^{L_0-1/24}&=&(1.q^0)+(1+1+1).q^1+(1+2+1).q^2+\ldots\nonumber \\
&=& (1+q+2q^2+\ldots)+2q(1+q+2q^2+\ldots)+2q^4(1+q+\ldots)+\ldots\nonumber \\
&=&\frac{1}{\eta(q)}\sum_{n\in\bZ}q^{n^2}
 \ .
\end{eqnarray}
In the basis where $J^1$ is diagonal, $g$ inverts the sign of ladder operators (or rather, $g$ anticommutes with $J^\pm$). Thus, inserting $g$ we obtain $g=(-1)^m$ and so
\begin{eqnarray}
tr\ gq^{L_0-1/24}&=&(1.q^0)+(-1+1-1).q^1+(-1+2-1).q^2+\ldots\nonumber \\
&=& (1+q+2q^2+\ldots)+2q(1+q+2q^2+\ldots)+2q^4(1+q+\ldots)+\ldots\nonumber \\
&=&\frac{1}{\eta(q)}\sum_{n\in\bZ}(-1)^n q^{n^2}
 \ .
\end{eqnarray}

With the deformation, the Hamiltonian is
shifted to $L_o=(m+\lambda/2)^2+(N-m^2)=N+\lambda m + \lambda^2/4$.
The $j=0$ contribution is
\begin{eqnarray}
tr\ q^{L_0-1/24}&=& q^{\lambda^2/4}
\{(1).q^0+(q^\lambda+1+q^{-\lambda}).q^1+(q^\lambda+2+q^{-\lambda}).q^2+\ldots
\} \nonumber \\
&=&\frac{1}{\eta(q)}\sum_{n\in\bZ}q^{n^2+\lambda^2/4}e^{\pi\lambda tn}
= \frac{1}{\eta (q)}\sum_{n\in Z} q^{(n-\lambda/2)^2} \ .
\end{eqnarray}
With the
$g$ insertion, we get
\begin{eqnarray}
tr\ gq^{L_0-1/24}&=&q^{\lambda^2/4}\{(1).q^0+(-q^\lambda+1-q^{-\lambda}).q^1
+(-q^\lambda+2-q^{-\lambda}).q^2+\ldots \}\nonumber \\
&=&\frac{1}{\eta(q)}\sum_{n\in\bZ}(-1)^{n}q^{n^2+\lambda^2/4}e^{\pi\lambda tn}
=\frac{1}{\eta(q)}\sum_{n\in\bZ}(-1)^{n}q^{(n-\lambda/2)^2} \ .
\end{eqnarray}
In the above expressions, we have $q=e^{-\pi t}$.
Poisson resummation and modular transformation then gives
\begin{eqnarray}
\tr \ q^{L_0-1/24}
&=& \frac{1}{\eta(q)}\sum_{n\in\bZ}e^{-\pi t (n -\lambda /2)^2} \nonumber \\
 &=&\frac{1}{\sqrt{2}\eta (\tilde q^{2})}\sum_{m\in\bZ}\tilde q^{m^2/2}
 e^{i\pi\lambda m} \ ,
\end{eqnarray}
where $\tilde q=e^{-2\pi /t}$. Similarly
\begin{eqnarray}
\tr \ gq^{L_0-1/24} &=&
\frac{1}{\eta(q)}\sum_{n\in\bZ}(-1)^{n}q^{(n-\lambda /2)^2} \nonumber \\
\mbox{}
&=& \frac{1}{\sqrt{2} \eta (\tilde q^2)} \sum_{m\in \bZ} \tilde q^{(m-1/2)^2/2}
e^{i\pi \lambda (m-1/2) } \ .
\end{eqnarray}


\begin{thebibliography}{10}
\newcommand{\arXiv}[1]{\href{http://xxx.lanl.gov/abs/#1}{{#1}}}

\bibitem{Callan}
  C.~G.~Callan, I.~R.~Klebanov, A.~W.~W.~Ludwig and J.~M.~Maldacena,
  ``{\it Exact solution of a boundary conformal field theory},''
  Nucl.\ Phys.\ B {\bf 422}, 417 (1994),
  \arXiv{hep-th/9402113}.

\bibitem{PT}
J.~Polchinski and L.~Thorlacius,
``{\it Free fermion representation of a boundary conformal field theory},''
Phys.\ Rev.\ D {\bf 50}, 622 (1994),
\arXiv{hep-th/9404008}.

\bibitem{Gutperle:2002ai}
  M.~Gutperle and A.~Strominger,
  JHEP {\bf 0204}, 018 (2002)
  [arXiv:hep-th/0202210].

\bibitem{Sen}
  A.~Sen,
  ``{\it Rolling tachyon},''
  JHEP {\bf 0204}, 048 (2002),
  \arXiv{hep-th/0203211}.

\bibitem{Larsen}
F.~Larsen, A.~Naqvi and S.~Terashima,
``{\it Rolling tachyons and decaying branes}'',
JHEP {\bf 0302}, 039 (2003),
\arXiv{hep-th/0212248}.

\bibitem{BHKN}
V.~Balasubramanian, S.~F. Hassan, E.~Keski-Vakkuri, and A.~Naqvi, ``{{\it A
  space-time orbifold: A toy model for a cosmological singularity,}}'' {\em
  Phys. Rev.} {\bf D67} (2003) 026003,
  \arXiv{hep-th/0202187}.

\bibitem{BKLNS}
  R.~Biswas, E.~Keski-Vakkuri, R.~G.~Leigh, S.~Nowling and E.~Sharpe,
  ``{\it The taming of closed time-like curves},''
  JHEP {\bf 0401}, 064 (2004),
  \arXiv{hep-th/0304241}.

\bibitem{uscosmo}
  S.~Kawai, E.~Keski-Vakkuri, R.~G.~Leigh and S.~Nowling,
  ``{\it Fractional S-branes on a Spacetime Orbifold},''
  to appear.

\bibitem{KKLN}
  S.~Kawai, E.~Keski-Vakkuri, R.~G.~Leigh and S.~Nowling,
  ``{\it Brane decay from the origin of time},''
  Phys.\ Rev.\ Lett.\ {\bf 96}, 031301 (2006),
  \arXiv{hep-th/0507163}.

\bibitem{Recknagel:1998ih}
A.~Recknagel and V.~Schomerus,
``{\it Boundary deformation theory and moduli spaces of D-branes},''
Nucl.\ Phys.\ B {\bf 545}, 233 (1999),
\arXiv{hep-th/9811237}.

\bibitem{AfflOsh}
M.~Oshikawa and I.~Affleck,
``{\it Boundary conformal field theory approach to the critical
two-dimensional Ising model with a defect line},''
Nucl.\ Phys.\ B {\bf 495}, 533 (1997),
\arXiv{cond-mat/9612187}.

\bibitem{Cardy}
  J.~L.~Cardy,
  ``{\it Boundary Conditions, Fusion Rules And The Verlinde Formula},''
  Nucl.\ Phys.\ B {\bf 324}, 581 (1989).

\bibitem{BYB}
  P.~Di Francesco, P.~Mathieu and D.~Senechal,
  ``{\it Conformal field theory}'',
   New York, USA: Springer (1997).

\bibitem{GR}
  M.~R.~Gaberdiel and A.~Recknagel,
  ``{\it Conformal boundary states for free bosons and fermions},''
  JHEP {\bf 0111}, 016 (2001),
  \arXiv{hep-th/0108238}.

\bibitem{LLM}
  N.~Lambert, H.~Liu and J.~Maldacena,
  ``{\it Closed strings from decaying D-branes},''
  \arXiv{hep-th/0303139}.

\bibitem{Gaiotto:2003rm}
  D.~Gaiotto, N.~Itzhaki and L.~Rastelli,
 ``{\it Closed strings as imaginary D-branes},''
  Nucl.\ Phys.\ B {\bf 688}, 70 (2004),
 \arXiv{hep-th/0304192}.

\bibitem{Karczmarek:2003xm}
  J.~L.~Karczmarek, H.~Liu, J.~Maldacena and A.~Strominger,
  ``{\it UV finite brane decay},''
  JHEP {\bf 0311}, 042 (2003),
  \arXiv{hep-th/0306132}.

\bibitem{Sen:2004zm}
A.~Sen,
``{\it Rolling tachyon boundary state, conserved charges and two dimensional  string
theory},''
JHEP {\bf 0405}, 076 (2004),
 \arXiv{hep-th/0402157}.




\bibitem{Affleck}
  I.~Affleck,
  ``{\it Conformal Field Theory Approach to the Kondo Effect},''
  Acta Phys.\ Polon.\ B {\bf 26}, 1869 (1995),
  \arXiv{cond-mat/9512099}.

\bibitem{KT}
  K.~R.~Kristjansson and L.~Thorlacius,
  ``{\it c = 1 boundary conformal field theory revisited},''
  Class.\ Quant.\ Grav.\  {\bf 21}, S1359 (2004),
  \arXiv{hep-th/0401003};
  ``{\it Correlation functions in a c = 1 boundary conformal field theory},''
  JHEP {\bf 0501}, 047 (2005),
  \arXiv{hep-th/0412175}.

\bibitem{GRW}
   M.~R.~Gaberdiel, A.~Recknagel and G.~M.~T.~Watts,
  ``{\it The conformal boundary states for SU(2) at level 1},''
  Nucl.\ Phys.\ B {\bf 626}, 344 (2002),
  \arXiv{hep-th/0108102};

  L.~S.~Tseng,
  ``{\it A note on c = 1 Virasoro boundary states and asymmetric shift  orbifolds},''
  JHEP {\bf 0204}, 051 (2002),
  \arXiv{hep-th/0201254};

 M.~R.~Gaberdiel and M.~Gutperle,
  ``{\it Remarks on the rolling tachyon BCFT},''
  JHEP {\bf 0502}, 051 (2005),
  \arXiv{hep-th/0410098}.


\end{thebibliography}

\providecommand{\href}[2]{#2}\begingroup\raggedright

\end{document}